\documentclass[conference]{IEEEtran}
\usepackage{amsfonts}
\usepackage{amsmath}
\usepackage{amssymb}
\usepackage{algorithm}
\usepackage{algorithmic}
\usepackage{cite}
\usepackage{hyperref}
\ifCLASSOPTIONcompsoc
  \usepackage[caption=false,font=normalsize,labelfont=sf,textfont=sf]{subfig}
\else
  \usepackage[caption=false,font=footnotesize]{subfig}
\fi
\ifCLASSINFOpdf   
   \usepackage[pdftex]{graphicx}     
      \graphicspath{{../pdf/}{../jpeg/}{./image/}}    
       \DeclareGraphicsExtensions{.pdf,.jpeg,.png,.jpg}   
 \else   
   \fi

\begin{document}
\title{Agent Based Intelligent Alert System for Smart-Phones}
\author{
\IEEEauthorblockN{Sandeep Venkatesh\IEEEauthorrefmark{1}, Shreyas Balakuntala\IEEEauthorrefmark{2}, Rajarajeswari S \IEEEauthorrefmark{3},\\ Nytika N Shetty\IEEEauthorrefmark{4}, Namratha Shetty\IEEEauthorrefmark{5} and Neha Sudhakar E\IEEEauthorrefmark{6}}
 \IEEEauthorblockA{Computer Science and Engineering\\ 
M S Ramaiah Institute of Technology, Bangalore, India 560-054\\}
\IEEEauthorblockA{\IEEEauthorrefmark{1}vsandeepnaidu@gmail.com, \IEEEauthorrefmark{2}shreyasbs1@yahoo.com, \IEEEauthorrefmark{3}gouthamraji8@gmail.com, \\\IEEEauthorrefmark{4}nytika.shetty@gmail.com, \IEEEauthorrefmark{5}namratha.7shetty@gmail.com, \IEEEauthorrefmark{6}nehasudhakar\_es@rediffmail.com}
} 
\maketitle

\begin{abstract}
The paper deals with the design of an agent which modifies and enhances the various alert systems in the smartphones. The actions of the agent includes sorting the notifications abiding to human thinking, helping the user to have a safe conversation, assisting in tracking back the  
reachability status of the caller when needed, conveying the user about the notifications in times of situations like drained battery and smartly alerting the user in situations like sleeping. The agent uses the information gathered from a survey, to modify the existing methods of alerts and produce alerts which abide by the human cognitive responses.

\end{abstract}

\section{Introduction}
Smart-phones have become an integral part of human lives. The alerts provided by the smart-phones, mostly follow a simple and traditional approach. But these alerts are in contrast with the human cognition. To start with, the sorting of notifications for calls, messages etc., take place in a time based sorting manner i.e., recent calls will be stacked on top of others. But the pattern in which the user calls back the callers whose calls he has missed, rarely follows this order. Secondly some of the profiles which help the phone to remain silent, in situations like sleeping, does it in a manner which is static, limited and also rarely considers the importance of the caller. Another problem to consider, is the smartphone's radiation hazards. The smart phone emits radiations, which are harmful to humans when exposed to a certain time period \cite{a}. The smart-phones lack a method which reduces this harmful effect, by warning the user. They also lack a method to track back the caller in case he is unreachable at the time of calling.This results in a toll on user's time, where he needs to call back repeatedly to check the status. The survey conducted on some volunteers, showed that the users feel the existing system should be modified for a betterment.
\section{Our Approach}
We propose an agent which maintains the information like user's classified set, probability of unsafe calls etc. in its knowledge base. The agent uses formulae derived from the survey which was conducted on volunteers to alert the user in a smart way. The agent does the tasks mentioned in "The System" subsection of this article. 
\section{Definitions}
\subsection{Alerts}
\subsubsection{Audible Alerts}
Audible alerts comprises of alerts like Audio alert for calls, messages, reminders, alarms etc.
\subsubsection{Physical Alerts}
Physical Alerts comprises of alert through the smartphone’s vibration alert hardware.
\subsubsection{Notification alerts}
Notification alerts comprises of alerts like missed call lists, message lists, battery status etc.
\subsubsection{Reminder Alerts}
Reminder alerts comprises of user-defined reminder alerts and machine produced reminder alerts, which we discuss in later sections of our paper.

\subsection{Context Awareness}
Context refers to the environment in which the smartphone is present. We say that the smartphone is \textit{context-aware}\cite{b} when the phone is fully/partially aware of its surroundings. A lot of work has been done in the area of context awareness using smart phones \cite{bb,c,d,e,f,g,h,i,j}.Our system uses the smartphone’s perception to determine its environment. Along with the user inputs, it uses the identity of connect-able devices (Wi-Fi hotspots, Audio Devices, Accessories), microphone’s perception, proximity sensor’s perception etc. to determine its environment like Home, Workspace, Driving, Outdoor etc. which covers the major requirement for our system to configure the alerts. 

\subsection{User Interactive}
The system uses the user input to configure the alerts to the maximum possible extent. The user input has the highest priority. In situations where user cannot give the input, the smartphone's context awareness is used to determine the situation.

\subsection{Classified Set}
The people whom the user communicates via his smart phones are classified in our system into 4 categories. The classification can be user interactive, machine learning based or a combination of both.
\subsubsection{Group A}
Group A comprises of people like parents, children, spouse etc. This group has the highest priority and is considered first in case of answering back in any situation.
\subsubsection{Group B}
Group B comprises of friends of high attachment, security staff, current project associates etc. This group has the second priority and is considered second in case of answering back in any situation.
\subsubsection{Group C}
Group C comprises of people like friends, people whose contacts are saved, distant relatives etc. This group has the third priority and is considered third in case of answering back in any situation.
\subsubsection{Group D}
Group D comprises of unknown numbers, promotional calls etc. This group has the least priority and is considered last in case of answering back in any situation.
\section{The System}
The problems suggested in the previous sections are handled by the agent. There are situations where the agent needs to know the situation of the user like whether he is sleeping, in a meeting etc. This can be perceived by the agent either by user input or by any other techniques like context-awareness. The agent's tasks is described in the following sections. 
\subsection{Caller Phone Tracker}
Any caller experiences one or more of the following problems often when calling a person\\
\subsubsection*{1}The phone of the person may be switched off.
\subsubsection*{2}The phone of the person may be out of reach.
\subsubsection*{3}The call may end abruptly due to some problem like drained battery.\\

When in such situation, the user is sometimes prone to make the calls repeatedly to check the status of other person's phone.This maybe due to the following conditions,
\subsubsection*{1}The person being unaware of the calls being made.
\subsubsection*{2}The person may be in a situation where he cannot make outgoing calls from his phone.\\

\begin{figure}[!t]
\centering
\includegraphics[width=9cm, height=7cm]{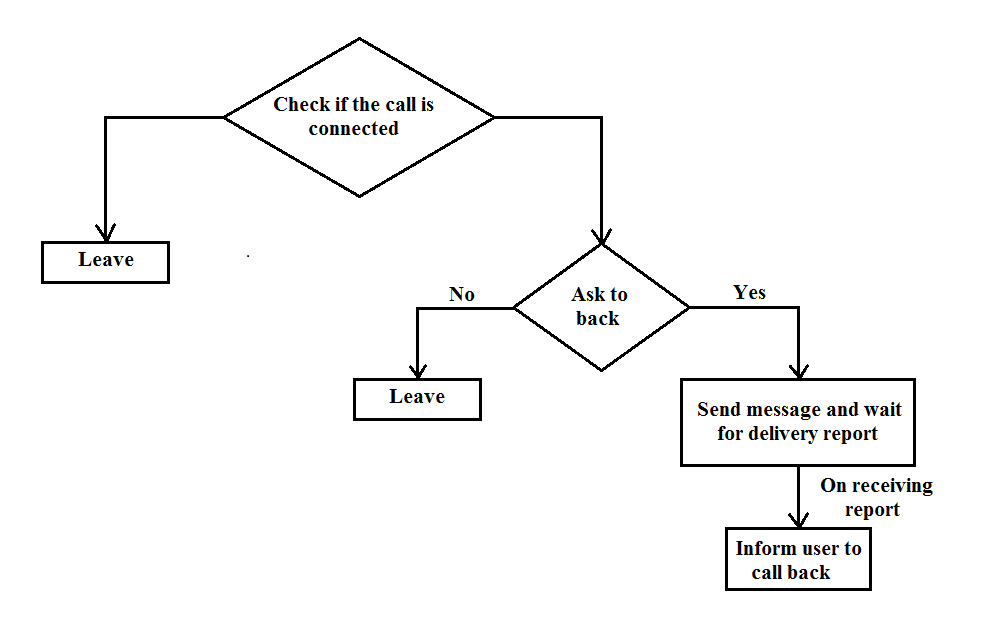}
\caption{A Flow chart which shows how the agent handles caller phone tracking}
\label{tracker}
\end{figure}

The agent takes care of the above situation by asking the user, whether the phone's status of the user wants to be tracked. The agent uses the delivery reports provided by the service provider present on the other person's phone. The agent goes to a state called "Do Nothing" where it remains idle when action is not needed. The flow chart~\ref{tracker} depicts the actions performed. Here the agent asks the user, whether to track back the
caller. Depending on the user's response the agent tracks the caller by sending a message to the caller and waits for the delivery report. It alerts the user on receiving a positive report.

\subsection{Smart Missed Call and Message Sorter}
It is a traditional approach, to display the calls we have missed and messages we have received in an order which is sorted purely on its time of arrival. On a survey conducted on 30 volunteers, of age group ranging from 20-40 years, it is inferred that the time based sorting, is of less use when it comes to the order in which they call back the callers. It was also inferred that the order which they wish to call back depends on the factors, like the classified group of the caller, sum total of the calls missed from the person, frequency and time distribution of the calls missed from the caller. We suggest a novel method to sort the calls missed and messages received from a caller considering the above factors. The agent follows the formula to sort the notifications. The highest value is given the highest priority in contacting back. 
$$[(Group Weight)*n]\over[T]$$
where Group A = Weight 4, Group B = Weight 3, Group C = Weight 2, Group D = Weight 1. T=The time difference between current time and time of the latest call from the user.

\subsection{Sleep mode alert}
Out of a survey conducted on people of age group ranging from 20-40 years, it was concluded that the mere alert mode of ringing/silent for all users is not effective. They also felt that the alerts of the calls while they are sleeping should be based on relative importance of the callers. The agent uses the suggested method to modify the existing alert method to provide a better and reasonable alert system. 
If the temporary importance is not set, it sets an alert like a ring or beep after the following number of calls:
$$[(4-Group Weight)*2]$$
Group A = Weight 4, Group B = Weight 3, Group C = Weight 2, Group D = Weight 1.
The temporary importance ignores the above factor and alerts the user for the first time. For group A, the value will be 0. This means the phone will alert the user the first time itself.

The same alert system can be extended to situations like meeting, working, studying etc.

\subsection{Alerts when the phone is switched off due to drained battery}

The smart phone users sometimes come across a situation where they are unable to use their phones. If the phone battery is very low, it is very important to preserve the leftover charge in its battery for emergency purposes. So answering calls which are of less importance is unnecessary. In such situations, the user may miss some notifications, calls and messages until he charges his smartphone and powers it on. Sometimes this period of waiting causes inconvenience for both the user and the caller. The agent takes care of this inconvenience by the following action,

It checks for the battery percentage, when it reaches below a critical point of 4\%, it saves all the notifications and call status, sorted in an order mentioned in the other sections and performs an action prioritised by the user.
The actions may be
\subsubsection*{1}To inform the caller that the battery is low.
\subsubsection*{2}To divert calls from Group 1 to a predefined number.
\subsubsection*{3}To send a text message to a predefined number.
\subsubsection*{4}To email the status to a predefined email id.

\subsection{Radiation warning} 
\begin{figure}[!t]
\includegraphics[width=10cm, height=12cm]{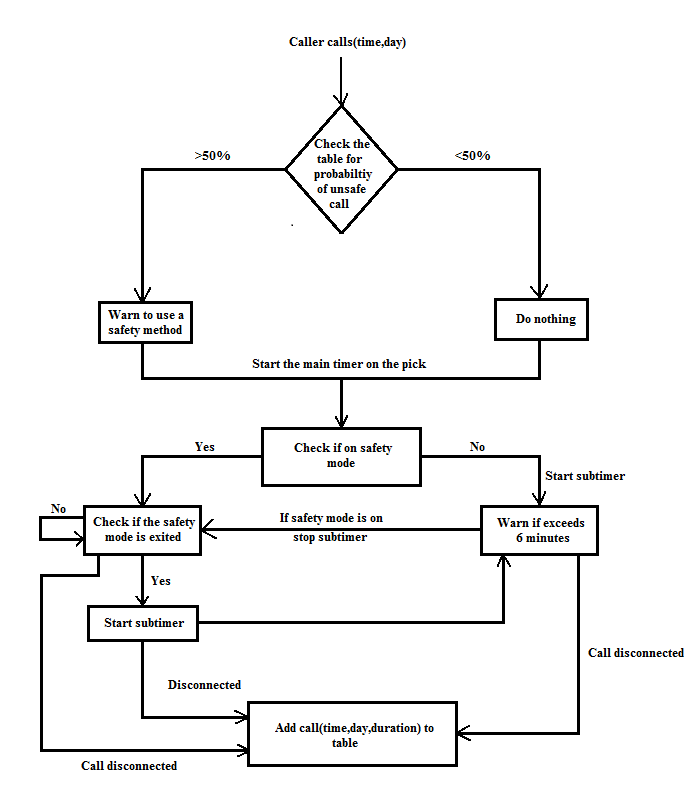}
\caption{A Flow chart which shows how the agent alerts for radiation hazards}
\label{radiation}
\end{figure}

The radiations emitted by a smart phone when exposed near human brain, leads to various health complications. It is said that the phone calls shouldn't last more than six minutes when held close to the ears. this effect is nullified when the conversations take place using utilities like speakerphone, headphones or any other connected devices. We suggest a method where the agent alerts the user to use the safety methods if the conversation exceeds the safe level. The agent also provides prior warning based on the previous conversations of the caller, maintained in the knowledge base. The terms used in the flow chart~\ref{radiation} are detailed below:
\subsubsection{Safety Method}
A safety method is when the user uses utilities like headphones, loudspeaker or any other connected devices to the conversation. This means that the phone is at a safer distance from the user's head, to nullify the hazards of radiation. The agent can use various methods like input from proximity sensors,connection status of utilities to determine whether or not the safety method is incorporated. Safety mode refers to incorporation of safety method.
\subsubsection{Probability of an unsafe call}
The agent maintains a table for each caller. The table contains the calls classified as safe and unsafe based on the previous conversations by the user. If the main timer of call had exceeded a time span of 6 minutes, the call is classified as unsafe. The probability is calculated from the table as a ratio of unsafe calls to total calls. The agent may also consider various factors like classified group, time of call, location etc. in a machine learning environment to calculate the probability of an unsafe call. 
\subsubsection{Main timer and sub timer}
The main timer is used to determine the total time length of conversation. It starts at the beginning of the call and ends at the termination. The subtimer starts when the user exits a safety mode or when not in a safety mode, at time of the beginning of the call. It terminates on entering a safety mode and restarts with a termination of safety mode.

\subsection{Alert through connected devices}
It is a common habit to leave the phone away in a place, when in places like home, office cabin etc. This sometimes makes the user to miss the calls/alerts by the phone. The agent reduces this problem by the method described further. The agent can infer the location based on methods like context aware computing or user interactive process. The agent then registers the devices connected like smart television, laptop computers. The user can register the device and required notifications like facebook messages, mail etc. with the agent. The agent sends the information of alerts, if not attended by the user, to the registered devices. This reduces the problem to a greater extent, as the user is at a high probability of using any of the connected devices at these places. 
\section{Results}
The survey conducted on 60 people of age ranging between 20-40, concluded the following
\subsubsection*{a}
\begin{figure}[!t]
\centering
\includegraphics[width=9cm, height=7cm]{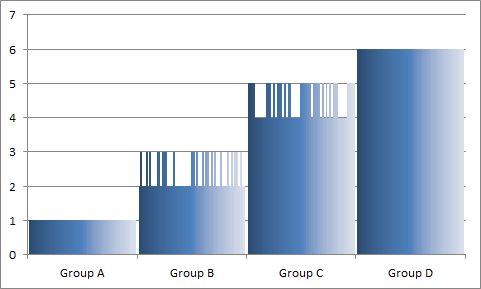}
\caption{A graph with groups on X-axis and when the user answers on Y-axis}
\label{survey}
\end{figure}

The phone calls are considered important, if called for a specific number of times based on the classified group. The calls are acceptably ignored, until it reaches the number specified in the survey, in situations like sleeping, meeting etc. This data is used in deriving the formulae used above.It is shown in Fig~\ref{survey} with Group on X axis and Number of calls in Y axis. Here 1 for Group A tells that the user considers the call as important for the first time and 6 in Group D denotes that the user considers the call as important if he gets it repeatedly for 6 times.
\subsubsection*{b} The sorting of notifications should be improved from the normal sorting method. 
\subsubsection*{c}Alerts during situations like sleeping, meeting, studying etc. should be based on importance of the call rather than a silent or ringing profile.
\subsubsection*{d} The smartphones, according to their knowledge lacks a method to warn about the radiation hazards and it would be better if the smartphones alert them about this. 

\section{Conclusion and Further works}
The paper suggests a design for an agent, which takes care of the problems suggested in the previous sections. The theoretical design can be implemented practically. Its function can be further improvised by considering more factors like surrounding sound, schedule of the user,mood of the user and subjecting the agent to machine learning techniques. The classification set can be modified and the members can be differentiated on factors like the call pattern, length of conversation, frequency of calls etc. 

\section*{Acknowledgement}
The authors would like to thank M.S.Ramaiah Institute of Technology for all the support and facilities, Prof.K.G.Srinivasa( Head, Dept of CSE, M.S.Ramaiah Institute of Technology) for his support and motivation. The authors would also like to thank the volunteers for their quality time and support.

\end{document}